\documentclass{revtex4-2}
\usepackage{eurosym}
\usepackage{amssymb}
\usepackage{amsmath}
\usepackage{graphicx}

\begin{document}

\title{The continuous spectrum of bound states in expulsive potentials: \\
self-trapping in the linear system }
\author{Hidetsugu Sakaguchi$^{1}$, Boris A. Malomed$^{2}$, Andreas C.
Aristotelous$^{3}$ and Efstathios G. Charalampidis$^{4}$}
\address{$^{1}$Interdisciplinary Graduate School of
Engineering Sciences, Kyushu University, Kasuga, Fukuoka 816-8580, Japan}
\address{$^{2}$Department of Physical Electronics, School of Electrical Engineering,
Faculty of Engineering, and Center for Light-Matter Interaction, Tel Aviv
University, P.O. Box 39040 Tel Aviv, Israel}
%\address{$^{2}$Instituto de Alta Investigaci\'{o}n, Universidad de Tarapac\'{a}, Casilla 7D,
%Arica, Chile}
\address{$^{3}$Department of Mathematics, The University of Akron, OH 44325,
USA}
\address{$^{4}$Department of Mathematics and Statistics and Computational
Science Research Center, San Diego State University, San Diego, CA 92182-7720, USA}

\begin{abstract}
On the contrary to the common intuition, which suggests that a steep
expulsive potential makes quantum states widely delocalized, we demonstrate
that one- and two-dimensional (1D and 2D) Schr\"{o}dinger equations, which
include expulsive potentials that are \emph{steeper than the quadratic}
ones, give rise to \emph{normalizable} eigenstates, which may be considered
as a manifestation of \emph{effective self-trapping in the linear system}.
These states constitute full continuous spectra in both the 1D and 2D cases.
In 1D, they are spatially even and odd eigenstates. The 2D states may carry
any value of the vorticity (alias magnetic quantum number). Asymptotic
expressions for wave functions of the 1D and 2D eigenstates, valid far from
the center, are derived analytically, demonstrating excellent agreement with
full numerical solutions. Special exact solutions for vortex states are
obtained in the 2D case. These findings suggest an extension of the concept
of bound states in the continuum, in quantum mechanics and paraxial
photonics. Gross-Pitaevskii equations are briefly considered as the
nonlinear extension of the 1D and 2D settings. In 1D, the cubic nonlinearity
slightly deforms the eigenstates, maintaining their stability
%On the other hand, the quintic self-focusing term, which occurs in the photonic version
% of the 1D model, initiates the dynamical collapse of states whose norm
%exceeds a critical value.

\end{abstract}

\maketitle

\section{Introduction and methods}

The separation of discrete and continuous spectra, which represent bound and
delocalized states, respectively, is an underlying principle of the
classical Sturm - Liouville theory \cite{SL1,SL2,SL3,SL4} and quantum
mechanics, whose mathematical framework is based on this theory \cite{LL,B2}%
. The fact that, in the paraxial approximation, the fundamental propagation
equation for optical waves is tantamount to the quantum-mechanical Schr\"{o}%
dinger equation makes the separation of discrete and continuous spectra an
equally important tenet of optics \cite{KA,Segev,LinOpt,Briggs,WangWei}.
This principle applies as well to other physical settings modeled by
equations of the Schr\"{o}dinger type \cite{Yang-book,Kevr}.

\begin{figure}[h]
\begin{center}
\includegraphics[height=5.0cm]{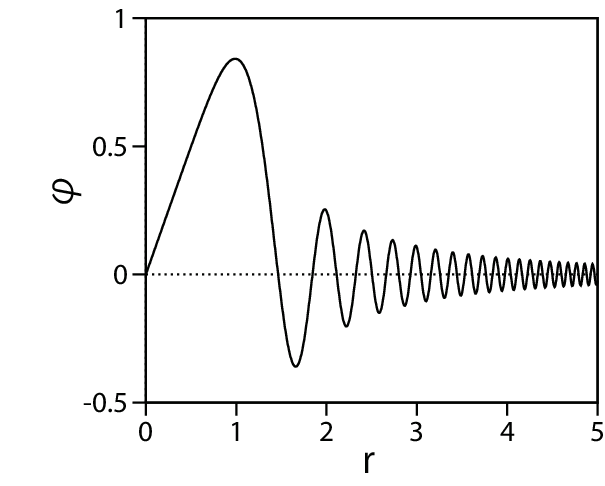}
\end{center}
\caption{The wave function $\protect\varphi _{\mathrm{vNW}}(r)$ of the von
Neumann-Wigner BIC\ state.}
\label{fig0}
\end{figure}

Nevertheless, there are well-known exceptions from the separation principle,
in the form of \textit{bound states embedded in the continuum} (often
designated by the BIC acronym). The first example was discovered in 1929 by
von Neumann and Wigner (vNW) \cite{bic1}, as a solution of the
three-dimensional (3D) Schr\"{o}dinger equation with the isotropic \emph{%
expulsive} potential (written here in a scaled form),%
\begin{equation}
U(r)=\frac{1}{r^{2}}-\frac{9}{2}r^{4}.  \label{vNW}
\end{equation}%
This potential supports a normalizable BIC stationary wave function,%
\begin{equation}
\varphi _{\mathrm{vNW}}(r)=\frac{1}{r^{2}}\sin \left( r^{3}\right)
\label{BIC}
\end{equation}%
(see Fig. \ref{fig0}), which corresponds to the zero energy eigenvalue.
Later, schemes were designed for the realization of quantum-mechanical BIC
states in semiconductor heterostructures \cite{bic3,bic4}. Then, much
interest was drawn to BIC modes in various photonic setups, where they offer
a considerable potential for applications \cite%
{BIC0,BIC1,BIC2,BIC7,BIC9,BIC6,BIC8,BIC10,BIC11,BIC12,BIC13}.

In the framework of the mean-field approximation, the linear Schr\"{o}dinger
equation for a single atom is closely related to its nonlinear version in
the form of the Gross-Pitaevskii equation (GPE), which includes the cubic
term accounting for the averaged effect of inter-atomic collisions in the
Bose-Einstein-condensate (BEC) phase of ultracold atomic gases \cite{Pit}.
In the optical realization, a model similar to GPE is based on the nonlinear
Schr\"{o}dinger equation (NLSE), which adds the Kerr self-focusing term to
the paraxial propagation equation \cite{KA}. Under special conditions,
NLSE-based optical models give rise to nonlinear states in the form of
\textit{embedded solitons}, which, similar to the BIC modes produced by the
linear Schr\"{o}dinger equation, may exist as peculiar localized states
\textit{embedded} in the continuous spectrum \cite{ES,ES5,ES2,ES3,ES4,ES6}.

Another counter-intuitive species of bound states is produced by a system of
two linearly coupled 1D or 2D equations, each one being either a linear or
nonlinear Schr\"{o}dinger equation, with the usual trapping
harmonic-oscillator (HO) potential in one equation, and the \emph{expulsive}
(anti-HO; alias inverted HO \cite{ODell}) potential in the other \cite{Nir}.
The 2D linear system for wave functions $u$ and $v$ , written in the polar
coordinates $\left( r,\theta \right) $, is%
\begin{eqnarray}
i\frac{\partial u}{\partial t}+\frac{1}{2}\left( \frac{\partial ^{2}}{%
\partial r^{2}}+\frac{1}{r}\frac{\partial }{\partial r}-\frac{1}{r^{2}}\frac{%
\partial ^{2}}{\partial \theta ^{2}}\right) u+\lambda v-\frac{1}{2}r^{2}u
&=&-\omega u,  \label{u2D} \\
i\frac{\partial v}{\partial t}+\frac{1}{2}\left( \frac{\partial ^{2}}{%
\partial r^{2}}+\frac{1}{r}\frac{\partial }{\partial r}-\frac{1}{r^{2}}\frac{%
\partial ^{2}}{\partial \theta ^{2}}\right) v+\lambda u+\frac{1}{2}\kappa
r^{2}v &=&0,  \label{v2D}
\end{eqnarray}%
where $\lambda $ is the coefficient of the linear coupling, $\kappa >0$ is
the relative strength of the anti-HO potential, while the HO strength in Eq.
(\ref{u2D}) is fixed to be $1$ by scaling, and $\omega $ represents a
possible eigenvalue mismatch between the coupled equations. It was found
\cite{Nir} that the system of Eqs. (\ref{u2D}) and (\ref{v2D}) admits
bound-state solutions, in the form of%
\begin{equation}
\left\{ u,v\right\} =\exp \left( -iEt+iS\theta \right) \left\{
U(r),V(r)\right\} ,  \label{vortex}
\end{equation}%
with integer vorticity $S=0,1,2,...$, real eigenvalue $E$ (alias the
chemical potential, in terms of BEC), which is a function of the system's
parameters $\lambda $, $\omega $, $\kappa $ and vorticity $S$ (as a special
example, see Eq.~\eqref{mu-exact} below), and stationary real wavefunctions
which are localized as usual eigenstates of the 2D HO, i.e., $\left\{
U(r),V(r)\right\} \sim r^{S}\exp \left( -r^{2}/2\right) $ at $r\rightarrow
\infty $, despite the presence of the expulsive potential in Eq. (\ref{v2D}%
). In particular, if the system's parameters are subject to constraint $%
\omega =(1/2)\left( 5+S-\lambda ^{2}\right) $, the bound state is found as
an exact solution,
\begin{gather}
u=U_{0}\left[ \left( \lambda ^{2}-1-S\right) +r^{2}\right] r^{S}\exp \left(
-iE_{\mathrm{exact}}t+iS\theta -\frac{r^{2}}{2}\right) ,  \label{u} \\
v=-2\lambda U_{0}r^{S}\exp \left( -iE_{\mathrm{exact}}t+iS\theta -\frac{r^{2}%
}{2}\right) ,  \label{v}
\end{gather}%
with the eigenvalue given by%
\begin{equation}
~E_{\mathrm{exact}}=\frac{1}{2}\left( \lambda ^{2}+1+S\right) ,
\label{mu-exact}
\end{equation}%
where $U_{0}$ is an arbitrary amplitude. These strongly localized states,
which exist with the single eigenvalue $E$, also represent BIC, as the same
system gives rise to the continuous spectrum of weakly delocalized
eigenstates, which are similar to the one given by Eq. (\ref{exact}) below.

The subject of the present work is to construct eigenstates of 1D and 2D
linear and (in a brief form) nonlinear Schr\"{o}dinger equations (alias
GPEs) with expulsive potentials, taken as the quadratic anti-HO term ($%
\gamma =1$ in Eq. (\ref{1D})) or steeper ones ($\gamma >1$; note that the
vNW example (\ref{vNW}) includes $\gamma =2$). The respective 1D GPE for
wave function $\psi \left( x,t\right) $ is introduced, in the scaled form,
as
\begin{equation}
i\frac{\partial \psi }{\partial t}=-\frac{1}{2}\frac{\partial ^{2}\psi }{%
\partial x^{2}}-\frac{1}{2}x^{2\gamma }\psi +g|\psi |^{2\sigma }\psi ,
\label{1D}
\end{equation}%
where the coefficient in front of the potential term is fixed by means of
scaling (cf. Eq. (\ref{u2D})), $g=-1,0,+1$ correspond to the self-focusing,
zero, or defocusing nonlinearity, and two physically relevant values, $%
\sigma =1$ and $2$, represent, respectively, the cubic and quintic
self-interaction. The cubic nonlinearity is the most generic one \cite{KA},
while the quintic self-focusing, which can be implemented \cite{Cid1} and
used \cite{Quiroga} under well-controlled conditions in optics \cite%
{Quiroga,Cid1}, is interesting, as it gives rise to the 1D variety \cite%
{Townes} of \textit{Townes solitons} \cite{Townes2,Townes3,Townes3}.
Eigenstates produced by Eq. (\ref{1D}) with real eigenvalue $E$ are looked
for in the usual form, $\psi \left( x,t\right) =\exp \left( -iEt\right)
\varphi (x)$, with stationary real eigenfunction $\varphi (x)$ satisfying
the equation
\begin{equation}
E\varphi =-\frac{1}{2}\frac{d^{2}\varphi }{dx^{2}}-\frac{1}{2}x^{2\gamma
}\varphi +g\varphi ^{2\sigma +1}.  \label{1Dphi}
\end{equation}

In the experiments with cold atoms, the expulsive anti-HO (quadratic) or
steeper potential, in its 1D and 2D forms alike, can be readily imposed by a
blue-detuned optical beam with the properly shaped transverse structure \cite%
{opt-beam2,opt-beam1,opt-beam3}, or a red-tuned one with intrinsic vorticity
\cite{vortex1,vortex2,vortex}. The recently developed technique of
programmable optical potentials \cite%
{Navon,programmable1,programmable2,programmable3} can also be used for this
purpose.
%Another possibility of the experimental realization of the 1D model
%represented by Eq. (\ref{1D}) refers to a gas of small polar molecules
%carrying a permanent electric dipole moment, cf. Refs. \cite{we,Trib}.
%Although this realization may seem too complex, one can consider the
%molecular gas loaded into a quasi-1D pipe-shaped optical trap, aligned with
%the $x$-axis, the dipoles being polarized in the positive and negative
%directions at $x>0$ and $x<0$, respectively, by a strong electric field
%generated by a uniformly charged sheet placed at $x=0$ perpendicular to the $%
%x$-axis. The expulsive potential $\sim 1/x^{2}$ ($\gamma =1$ in Eq. (\ref{1D}%
%)) is then imposed by a point-like charge additionally placed at $x=0$, with
%the sign opposite to that that of the polarizing sheet.

One may intuitively expect that the strongly expulsive potential in Eq.~(\ref%
{1D}) leads to strong delocalization of the respective eigenstates.
Nevertheless, we demonstrate, analytically and numerically, that the
intuition is misleading: the 1D and 2D linear Schr\"{o}dinger equations with
the anti-HO (quadratic) potential, $\gamma =1$, give rise to weakly
delocalized states, whose norm,%
\begin{equation}
N_{\mathrm{1D}}=\int_{-\infty }^{+\infty }\varphi ^{2}(x)dx,  \label{N1D}
\end{equation}%
diverges logarithmically with the increase of the size of the solution
domain (see Eq. (\ref{N}) below), while all steeper expulsive potentials,
corresponding to $\gamma >1$, produce \emph{effectively localized
(self-trapped) bound states} with a convergent norm, which form a continuous
spectrum ($-\infty <E<+\infty $), This situation seems unusual in quantum
mechanics and BEC theory (as sort of a \emph{quantum anomaly} \cite%
{we,ODell,Trib,Maxim,anomaly}) -- in particular, because self-trapping is
usually considered as the feature which may only be induced by the
nonlinearity, while here it takes place in the linear setting.

While the most essential results are reported here in the framework of the
1D and 2D linear Schr\"{o}dinger equations, some results for the 1D NLSE are
reported, in a brief form, too. It is demonstrated that numerically found 1D
bound states remain stable or develop parity-breaking instability, below or
above a critical value of the norm $N_{\mathrm{1D}}$, respectively, under
the action of the cubic self-focusing term ($\sigma =1$, $g=-1$) in Eq. (\ref%
{1D}) with the quartic expulsive potential ($\gamma =2$).
%The quintic self-focusing term ($\sigma =2$ in
%Eq. (\ref{1D}), with $g=-1$) gives rise to the collapse of the 1D bound
%states if their norm exceeds a certain critical value, while they are stable
%below the critical level. This situation is typical for the Townes-soliton
%phenomenology \cite{Berge,Kuz,Fibich,Ben-Li}.

Analytical and numerical results for the bound states in the 1D\ linear Schr%
\"{o}dinger equation are reported in subsection 2A. This is followed, in
subsection 2B, by a brief consideration of nonlinear bound states and their
stability in the above-mentioned case, with parameters $\sigma =1$, $g=-1$,
and $\gamma =2$ in Eq. (\ref{1D}),%and quintic
Findings for the 2D bound states, including vortex ones, which carry the
angular momentum, are presented in subsection 2C (chiefly, for the linear
Schr\"{o}dinger equation). In particular, the consideration of the 2D linear
Schr\"{o}dinger equation with the expulsive potential produces \emph{exact
solutions} for the vortex states, under a special constraint imposed on the
model's parameters. The paper is concluded by Section 3.

\section{Results}

\subsection{One-dimensional linear bound states}

The starting point of the analysis is the linear version ($g=0$) of the 1D
stationary equation (\ref{1Dphi}). While the equation is not analytically
solvable, it is straightforward to construct its asymptotic solution, valid
for sufficiently large values of $|x|$, in the form of a combination of
rapidly oscillating cosines and sines, with amplitudes expanded in
appropriate powers of $|x|^{-1}$. The analytical calculation is similar to
the commonly known one which produces the asymptotic approximation for the
Airy function \cite{LL}. Thus, the asymptotic approximation yields the
result which is valid for $\gamma \neq 1$:%
\begin{equation}
\varphi _{\mathrm{asympt}}^{\mathrm{(1D)}}(x;\gamma \neq 1;E)=\varphi
_{0}|x|^{-\gamma /2}\cos \left( \frac{|x|^{\gamma +1}}{\gamma +1}-\chi
_{0}\right) +\frac{E\varphi _{0}}{\gamma -1}|x|^{-3\gamma /2+1}\sin \left(
\frac{|x|^{\gamma +1}}{\gamma +1}-\chi _{0}\right) ,  \label{asympt}
\end{equation}%
where the amplitude $\varphi _{0}$ and phase shift $\chi _{0}$ are arbitrary
constants, in terms of the asymptotic solution. Note that setting $\gamma
=1/2$ in the first term in Eq. (\ref{asympt}) reproduces the above-mentioned
asymptotic approximation for the Airy function. The approximation (\ref%
{asympt}) retains the first two terms of the expansion, the third-order
correction ($\mathrm{TOC}$) being%
\begin{equation}
\mathrm{TOC}\left( \varphi _{\mathrm{asympt}}^{\mathrm{(1D)}}(x);\gamma \neq
1\right) =\frac{1}{8}\frac{\gamma \left( \gamma +2\right) }{\gamma +1}%
\varphi _{0}|x|^{-3\gamma /2-1}\sin \left( \frac{|x|^{\gamma +1}}{\gamma +1}%
-\chi _{0}\right) .  \label{TOC}
\end{equation}

An obvious property of the wave function (\ref{asympt}) is the \emph{%
convergence} of its integral norm, $N=\int_{-\infty }^{+\infty }\varphi
^{2}(x)dx$ at $|x|\rightarrow \infty $, for $\gamma >1$ and all eigenvalues $%
-\infty <E<+\infty $ (the full continuous spectrum). Thus, we arrive at a
simple but counter-intuitive conclusion: any expulsive potential which is
steeper than the quadratic (alias anti-HO) one, with $\gamma >1$, produces
\emph{normalizable} bound states, which populate the full continuous
spectrum. Qualitatively, this conclusion can be explained by noting that a
classical counterpart of the quantum particle would be rolling down the
steep potential hill with the rapidly growing acceleration, which gives rise
to a strongly oscillating phase in the respective wave function, the latter
feature leading to the \emph{effective self-trapping}
(localization/normalizability) of the eigenstate in the \emph{linear system}%
, unlike the commonly known mechanism of self-trapping (creation of
solitons) in various nonlinear models \cite{KA}.\-

Note that all inflexion points ($d^{2}\varphi /dx^{2}=0$) of the stationary
wave function produced by Eq. (\ref{1Dphi}) with $g=0$ and $E\geq 0$
coincide with zero-crossing ones, $\varphi (x)=0$. In the case of $E<0$,
there are two additional inflexion points, \textit{viz}., $x=\pm \left(
-2E\right) ^{1/(2\gamma )}$.

The situation is different in the case of the anti-HO potential, with $%
\gamma =1$ in Eq. (\ref{1D}). In this case, approximation (\ref{asympt}) is
replaced by the following one, which includes a logarithmic correction to
the phase of the oscillatory wave function:%
\begin{equation}
\varphi _{\mathrm{asympt}}^{\mathrm{(1D)}}(x;\gamma =1)=\varphi
_{0}|x|^{-1/2}\cos \left( \frac{x^{2}}{2}+E\ln \left( \frac{|x|}{l}\right)
\right) ,  \label{gamma=1}
\end{equation}%
where $l$ is a characteristic scale of the inner core of the wave function
(an exact wave function of the 1D anti-HO potential can be formally
expressed in terms of the special parabolic-cylinder function \cite{ODell}).
It follows from expression (\ref{gamma=1}) that the respective normalization
integral (\ref{N1D}) slowly diverges,
\begin{equation}
N_{\mathrm{1D}}\simeq \varphi _{0}^{2}\ln \left( L/l\right) ,  \label{N}
\end{equation}%
$2L$ being the size of the integration domain.

\begin{figure}[h]
\begin{center}
\includegraphics[height=5.0cm]{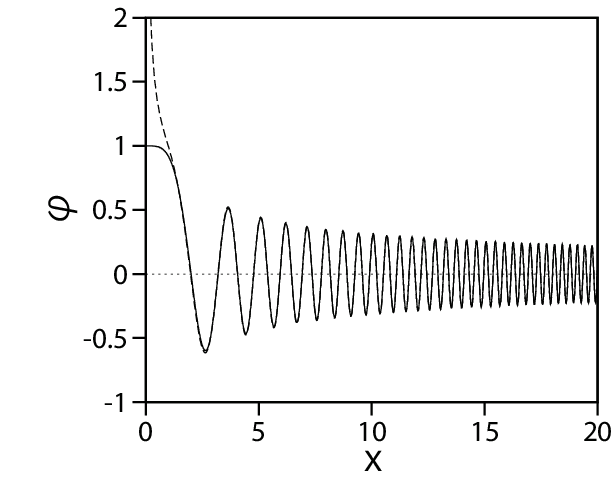}
\end{center}
\caption{The continuous curve: the numerically found spatially even solution
of Eq. (\protect\ref{1Dphi}) with $g=0$, $\protect\gamma =1$ (the quadratic
expulsive potential) and $E=0$. The dashed curve: the asymptotic
approximation (\protect\ref{asympt}) for the same solution, with fitting
constants $\protect\chi _{0}=\protect\pi /8$ and $\protect\varphi _{0}=1$.}
\label{fig1}
\end{figure}

The wave function of the 1D linear eigenstate, produced by a numerical
solution of Eq. (\ref{1Dphi}) with $g=0$, $\gamma =1$ (the anti-HO
potential) and $E=0$, and its comparison to the asymptotic approximation (%
\ref{asympt}), are plotted in Fig. \ref{fig1}, with fitting constants $\chi
_{0}=\pi /8$ and $\varphi _{0}=1$ in Eq. (\ref{asympt}) (in fact, $\varphi
_{0}$ is an arbitrarily constant for the linear solution). The respective
numerical solution of the second-order ordinary differential equation (\ref%
{1Dphi}) was produced by means of the Runge-Kutta method, with the initial
conditions
\begin{equation}
\varphi (x=0)=1,\frac{d\varphi }{dx}(x=0)=0  \label{initial}
\end{equation}%
(which means that even solutions, with $\varphi (-x)=\varphi (x)$, were
sought for). The same numerical method was used to obtain various
eigenstates in what follows below. It is clearly observed that the
analytical asymptotic approximation provides high accuracy, in comparison to
its numerical counterpart.

In Fig. \ref{fig1} it is not easy to display slowly decaying tails of the
wave function, as the rapidly accelerating phase oscillations in Eq. (\ref%
{asympt}) imply that the accurate numerical solution must take into regard
spatial harmonics of very high orders. On the other hand, the asymptotic
expression (\ref{asympt}) yields the tails in the virtually exact analytical
form.

\begin{figure}[h]
\begin{center}
\includegraphics[height=5.0cm]{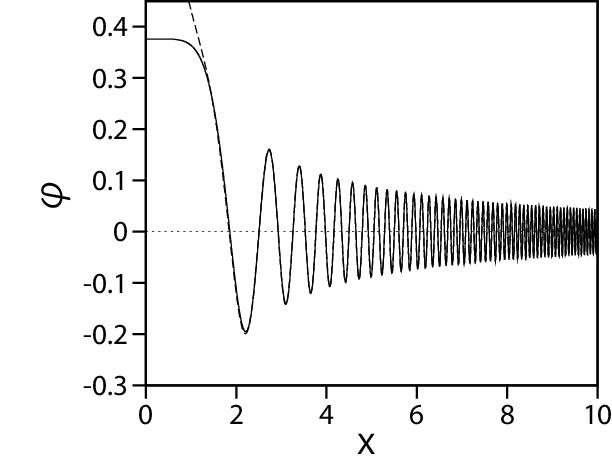}
\end{center}
\caption{The continuous curve: the numerically found spatially even solution
of Eq. (\protect\ref{1Dphi}) with $g=0$, $\protect\gamma =2$ (the quartic
expulsive potential) and $E=0$. The dashed curve: the asymptotic
approximation (\protect\ref{asympt}) for the same case, with fitting
constants $\protect\chi _{0}=\protect\pi /6$ and $\protect\varphi _{0}=0.44$%
. }
\label{fig2}
\end{figure}

A typical example of the \emph{normalizable} (effectively self-trapped) 1D
bound state for $\gamma =2$ (the quartic expulsive potential), produced by
the numerical solution of Eq. (\ref{1Dphi}), and its comparison to the
asymptotic approximation (\ref{asympt}), is displayed in Fig. \ref{fig2}. In
this case too, the asymptotic approximation demonstrates remarkably high
accuracy.

%\noindent\textit{Note:} For visualization purposes, the wavefunctions in
%several numerical figures were vertically rescaled. Accordingly, the plotted
%amplitudes do not represent the original normalization of the stationary
%solutions, and any amplitude-related fitting parameters (such as $%
%\varphi_{0} $) correspond to the rescaled profiles shown in the plots.

\begin{figure}[h]
\begin{center}
\includegraphics[height=5.0cm]{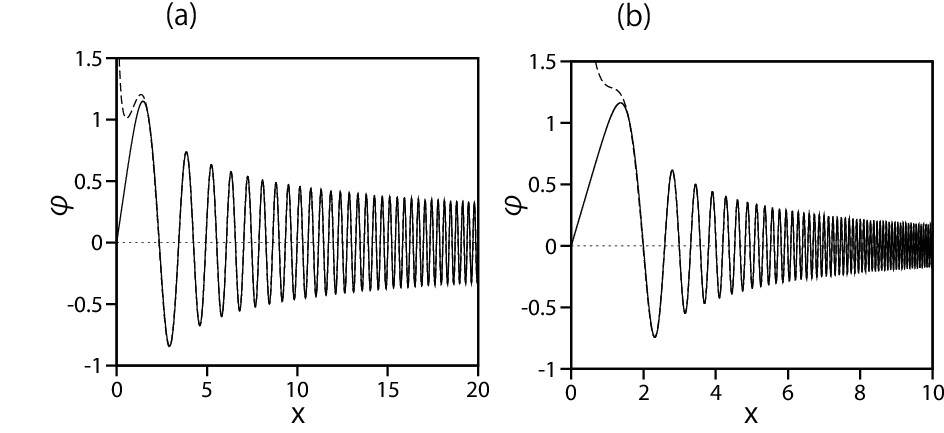}
\end{center}
\caption{In panels (a) and (b), the continuous curves represent numerically
found spatially odd (dipole-mode) solutions of Eq. (\protect\ref{1Dphi})
with $g=0$ and $E=0$, for $\protect\gamma =1$ and $2$, respectively (cf.
their spatially even (fundamental) counterparts displayed in Figs. \protect
\ref{fig1} and \protect\ref{fig2}, resectively). The dashed curves represent
the corresponding asymptotic approximation (\protect\ref{asympt}) with
fitting constants $\protect\chi _{0}=3\protect\pi /8$, $\protect\varphi %
_{0}=1.45$ in (a), and $\protect\chi _{0}=\protect\pi /3$, $\protect\varphi %
_{0}=1.72$ in (b).}
\label{fig3}
\end{figure}

For given $E$, the asymptotic expressions (\ref{asympt}) and (\ref{TOC}) may
belong, at least, to two different global solutions (full eigenstates),
\textit{viz}., the \textit{fundamental\ }spatially even one, with $\varphi
(-x)=\varphi (x)$ (alias the ground state, for given $E$), and the \textit{%
dipole} spatially odd eigenstate, with $\varphi (-x)=-\varphi (x)$ (alias
the first excited state). To illustrate this possibility for the same cases
as considered above for the fundamental solutions, \textit{viz}., $\gamma =1$
and $\gamma =2$, the corresponding numerically found dipole eigenstates are
plotted in Figs. \ref{fig3}(a) and (b), respectively. To this end, Eq.~(\ref%
{1Dphi}) was numerically solved again by means of the Runge-Kutta method,
with the initial conditions
\begin{equation}
\varphi (x=0)=0,~\frac{d\varphi }{dx}\left( x=0\right) =1,  \label{initial2}
\end{equation}%
cf. Eq. (\ref{initial}). It is plausible that Eq. (\ref{1Dphi}) may produce
additional spatially even and odd solutions, which correspond to
higher-order excited states, in terms of quantum mechanics. This possibility
will be considered elsewhere.

\begin{figure}[h]
\begin{center}
\includegraphics[height=5.0cm]{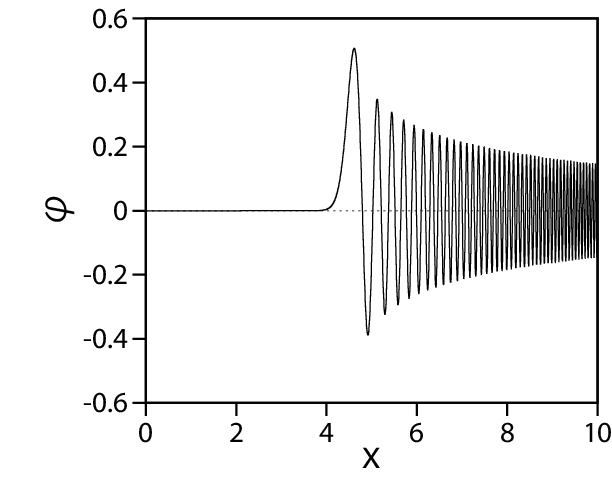}
\end{center}
\caption{The spatially even eigenstate produced by the numerical solution of
Eq. (\protect\ref{1Dphi}) with $g=0$, $\protect\gamma =2$, and $E=-200$.}
\label{fig4}
\end{figure}

\begin{figure}[h]
\begin{center}
\includegraphics[height=5.0cm]{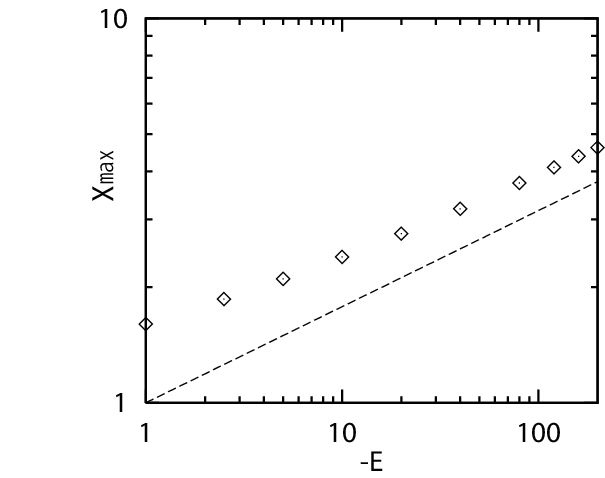}
\end{center}
\caption{The numerically found dependence of the coordinate $x_{\max }$, at
which the wave function attains its maximum, on the eigenvalue $E<0$, for $%
g=0$ and $\protect\gamma =2$. The dependence is plotted on the log-log
scale, with the straight dashed line representing the approximate relation (%
\protect\ref{1/4}).}
\label{fig5}
\end{figure}

While the above examples display eigenstates found with the zero eigenvalue (%
$E=0$), Fig. \ref{fig4} presents a typical eigenstate for a very large
negative eigenvalue, \textit{viz}., $E=-200$, in the case of $\gamma =2$
(the quartic expulsive potential). It is seen that the large value of $-E$
suppresses the wave function in the core area of the eigenstate.
%As the energy decreases, the wave function goes away from the potential maximum
%around $x=0$. As $E\rightarrow -\infty $, $x_{\max }\rightarrow \infty $ and
%the wave function approaches the ground state.
At the inner edge of the core, the wave function quickly grows towards its
largest value, attained at a point with $|x|=x_{\max }$, as $\varphi (x)\sim
\exp \left( \sqrt{-2E}\left( x-x_{\max }\right) \right) $, in agreement with
Eq. (\ref{1Dphi}). This is followed by the gradual decay of the rapidly
oscillating tail, in agreement with Eq. (\ref{asympt}). The dependence of $%
x_{\max }$ on $|E|$ may be predicted, in a crude approximation, by equating
the two respective large terms in Eq. (\ref{1Dphi}), i.e., $-E\varphi \sim
(1/2)x_{\max }^{4}\varphi $ (recall we here set $\gamma =2$), which yields,
in the logarithmic approximation,
\begin{equation}
\ln \left( x_{\max }\right) \simeq (1/4)\ln \left( -E\right) .  \label{1/4}
\end{equation}%
This simple relation is compared to the numerical data in Fig. \ref{fig5}.

In the eigenstates corresponding to\ very large positive eigenvalues, the
core area, $|x|\lesssim E^{1/4}$, is not (nearly) empty, unlike Fig. \ref%
{fig5}. Instead, it is filled by a nearly uniform standing wave, $\varphi
(x)\approx \varphi (x=0)\cos \left( \sqrt{2E}x\right) $ (not shown here in
detail).

\subsection{Stable and unstable bound states in the 1D cubic nonlinear Schr%
\"{o}dinger equation (NLSE) with the expulsive potential}

While the stability of the existing eigenstates in the framework of the
linear Schr\"{o}dinger equation is obvious, it is a nontrivial problem in
the case of the NLSE. The numerical solution of the stationary version of
Eq. (\ref{1D}) with the cubic nonlinearity, i.e., $\sigma =1$ and $g=\pm 1$,
has demonstrated the existence of the corresponding bound states, with
profiles somewhat deformed by the cubic term. Simulations of the perturbed
evolution of the bound states have demonstrated that they remain stable in
some parameter region, and become unstable in another, as briefly
demonstrated below.

\begin{figure}[h]
\begin{center}
\includegraphics[height=5.0cm]{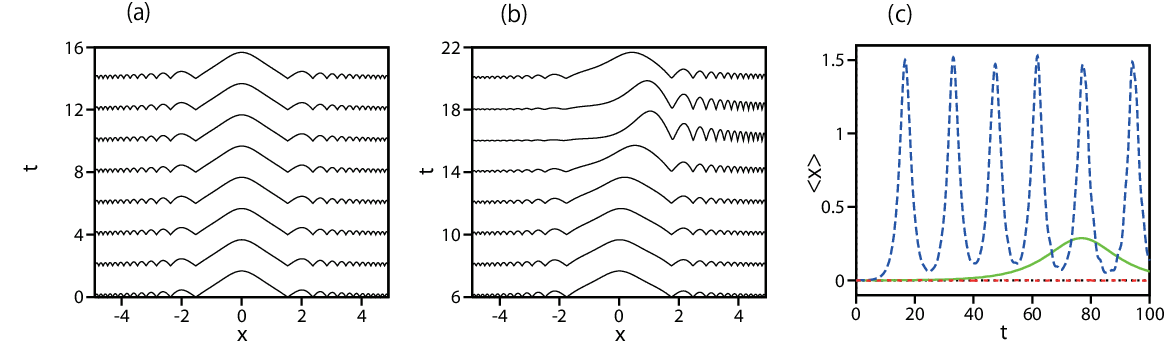}
\end{center}
\caption{(a) The stable evolution of the spatially even eigenstate obtained
as the numerical solution of Eq. (\protect\ref{1Dphi}) with $\protect\gamma %
=2$ , $g=-1$, $\protect\sigma =1$ (the cubic self-focusing nonlinearity),
eigenvalue $E=-0.9$, and norm $N_{\mathrm{1D}}=1.73$. Eight snapshots of $|%
\protect\psi (x,t)|$ at $t=0,2,\cdots ,14$ are plotted. (b) Snapshots of $|%
\protect\psi \left( x,t\right) |$ at $t=6,8,\cdots ,20$ illustrate the onset
of the spontaneous parity-breaking instability of the even eigenstate with $%
E=-1.1$ and $N_{\mathrm{1D}}=3.07$. (c) The center-of-mass coordinate (%
\protect\ref{com}) as a function of. time for the perturbed evolution of the
eigenstates with $E=-0.9$ (the dotted red line), $E=-1$ (the solid green
line), and $E=-1.1$ (the dashed blue line).}
\label{fig6}
\end{figure}

A typical example of the stable evolution of the bound state for $g=-1$ (the
self-focusing cubic term), $\gamma =2$ (the quartic expulsive potential),
and $E=-0.9$ is presented in Fig.\ \ref{fig6}(a). In this case, the
stationary even eigenstate (with $\varphi (-x)=\varphi (x)$) was obtained,
as the solution of the corresponding ordinary differential equation (\ref%
{1Dphi}), by dint of the Runge-Kutta method with the initial conditions $%
\varphi (x=0)=1.381$ and $d\varphi /dx(x=0)=0$, cf. Eq. (\ref{initial}). The
stationary wave function vanishes at edges of the solution domain, $%
|x|=L=4.8981$. The respective norm of the eigenstate, $\int_{-L}^{+L}\varphi
^{2}(x)dx$ (cf. Eq. (\ref{N1D})) is
\begin{equation}
N_{\mathrm{1D}}=1.73.  \label{1.73}
\end{equation}

To confirm the stability of this eigenstate, we have performed simulations
of its perturbed evolution in the framework of the full equation (\ref{1D}),
using the split-step Fourier method in the domain $|x|\leq L$, with zero
boundary conditions, $\psi (x=\pm L)=0$. The number of Fourier modes was $%
8192$, and the marching timestep was $\Delta t=10^{-5}$. As the local
wavenumber of the oscillating tail increases with $|x|$, more Fourier modes
are necessary for the simulations of a larger system. Here, we present the
results of the simulations performed in the domain of size $2L=9.7962$,
which maintain sufficiently high numerical accuracy. As the initial
condition, we took
\begin{equation}
\psi (x,t=0)=\varphi (x)+0.001\sin (2\pi x/L),  \label{t=0}
\end{equation}%
where the second term introduces a small spatially odd perturbation, which
breaks the parity of the nonlinear even eigenstate $\varphi (x)$. The aim
was, in particular, to test the stability of the eigenstate against the
spatial-symmetry-breaking perturbations. Eight snapshots of instantaneous
profiles $|\psi (x,t)|$, produced by the simulations at $t=0,2,\cdots ,14$,
are plotted in Fig.~\ref{fig6}(a). The presented results clearly confirm the
stability of the nonlinear eigenstate.

A symmetry-breaking instability of the eigenstate occurs for larger absolute
values of eigenvalue $E$, An example is presented in Fig. \ref{fig6}(b) for $%
g=-1$ and $E=-1.1$. The corresponding stationary eigenstate is produced by
the numerical solution of Eq. (\ref{1Dphi}) with the initial conditions $%
\varphi (x=0)=1.381$ and $d\varphi /dx\left( x=0\right) =0$, in the domain $%
0<x<L=4.9253$. The norm of the eigenstate is $N_{\mathrm{1D}}=3.07$, cf. the
above value (\ref{1.73}). Then, the perturbed evolution of the eigenstate
was simulated with the input taken as per Eq. (\ref{t=0}) (as above). Figure
~\ref{fig6}(b) shows snapshots of $|\psi (x,t)|$ at $t=6,8,\cdots ,20$. It
is seen that, unlike the stable evolution displayed in Fig. \ref{fig6}(a),
the odd perturbation grows, so that the central-peak's position moves to the
right and then returns to the origin, due to the interaction with the the
oscillatory tail.

The motion of the center of mass of the solutions,%
\begin{equation}
\langle x\rangle =\frac{1}{N_{\mathrm{1D}}}\int_{-L}^{+L}x|\psi (x)|^{2}dx,
\label{com}
\end{equation}%
is plotted in Fig. \ref{fig6}(c) for the above-mentioned perturbed
eigenstates with $E=-0.9$ (the dotted red line), $E=-1.1$ (the dashed blue
line), and also for the intermediate case, with $E=-1.0$, with $L=4.9116$
and $N_{\mathrm{1D}}=2.86$ (the solid green line). It is concluded that the
parity-breaking instability of the spatially even nonlinear eigenstates
occurs at $E\approx -1$, a natural conclusion being that the self-focusing
nonlinearity leads to the destabilization of the bound states above a
critical value of the norm. In particular, the dashed blue trajectory in
Fig. \ref{fig6}(c) demonstrates that cycle of the instability development,
which is displayed in Fig. \ref{fig6}(b), repeats with approximate
periodicity. Detailed studies of the stability of the nonlinear eigenstates,
and production of the respective stability charts for the systems with of
the cubic and quintic nonlinear terms, will be a subject of another work.

\begin{figure}[h]
\begin{center}
\includegraphics[height=5.0cm]{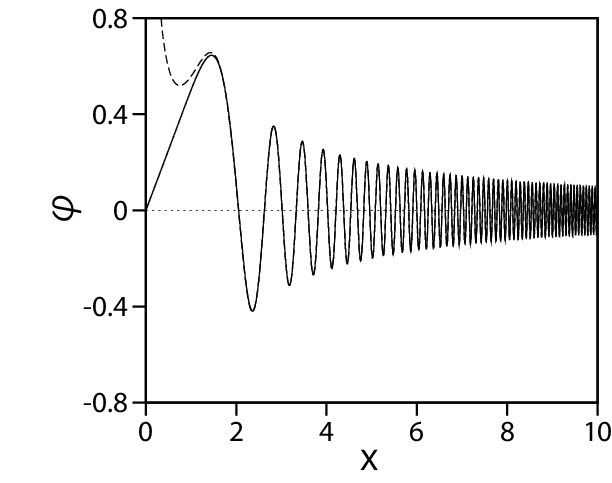}
\end{center}
\caption{The stable spatially odd eigenstate produced by the numerical
solution of Eq. (\protect\ref{1Dphi}) with $g=+1$, $E=0$, and $\protect%
\gamma =2$, cf. its counterpart plotted in Fig. \protect\ref{fig3}(b)
produced for the same parameters by the linearized equation (with $g=0$).
The norm of the eigenstate is $N=1.23$. The dashed curve represents the
corresponding asymptotic approximation (\protect\ref{asympt}) with fitting
constants $\protect\chi _{0}=5\protect\pi /12$ and $\protect\varphi %
_{0}=0.995$.}
\label{fig7}
\end{figure}

The 1D spatially odd (dipole) eigenstates are also weakly deformed by the
cubic nonlinearity, and may remain stable, depending on the parameters. An
example is displayed in Fig. \ref{fig7} for $\gamma =2$, $\sigma =1$, $g=+1$
(the self-defocusing cubic term) and $E=0$, i.e., the same parameters as
those (except for $g=0$) of the dipole-mode eigenstate shown in Fig. \ref%
{fig3}(b), which was produced by the linear version of Eq. (\ref{1Dphi}).
Note that both eigenstates are very well approximated by the asymptotic
expression (\ref{asympt}) with appropriate values of the fitting constants,
\textit{viz}., $\chi _{0}=\pi /3$ for $g=0$ and $\chi _{0}=5\pi /12$ for $%
g=+1$.

\subsection{Two-dimensional bound states}

The natural 2D version of Eq. (\ref{1D}) can also be realized experimentally
as the GPE for ultracold atoms, under the action of the isotropic expulsive
optical potential, induced by the blue-detuned laser beam, or the
red-detuned one with intrinsic vorticity. In terms of the polar coordinates $%
\left( r,\theta \right) $, the 2D GPE is written as
\begin{equation}
i\frac{\partial \psi }{\partial t}=-\frac{1}{2}\left( \frac{\partial
^{2}\psi }{\partial r^{2}}+\frac{1}{r}\frac{\partial \psi }{\partial r}+%
\frac{1}{r^{2}}\frac{\partial ^{2}\psi }{\partial \theta ^{2}}\right) -\frac{%
1}{2}r^{2\gamma }\psi +g|\psi |^{2}\psi ,  \label{2DNLS}
\end{equation}%
with $\gamma \geq 1$ (here, only the cubic nonlinear term is considered, if
any). The stationary version of Eq. (\ref{2DNLS}) with integer vorticity $S$
(alias magnetic quantum number) and eigenvalue $E$ (the chemical potential,
in terms of the BEC), is produced by the substitution
\begin{equation}
\psi =\exp \left( -iEt+iS\theta \right) \varphi (r),  \label{psiphi}
\end{equation}%
where real function $\varphi (r)$ obeys the equation%
\begin{equation}
E\varphi =-\frac{1}{2}\left( \frac{d^{2}\varphi }{dr^{2}}+\frac{1}{r}\frac{%
d\varphi }{dr}-\frac{S^{2}}{r^{2}}\varphi \right) -\frac{1}{2}r^{2\gamma
}\varphi +g\varphi ^{3}.  \label{g}
\end{equation}

Similar to Eq. (\ref{asympt}) in the 1D case, it is straightforward to
construct the asymptotic form of the tail of the solution to Eq. (\ref{g})
at $r\rightarrow \infty $, in the case of $\gamma \neq 1$. The first two
terms of the asymptotic expansion are
\begin{equation}
\varphi _{\mathrm{asympt}}^{\mathrm{(2D)}}(r)=\varphi _{0}r^{-(\gamma
+1)/2}\cos \left( \frac{r^{\gamma +1}}{\gamma +1}-\chi _{0}\right) +\frac{%
E\varphi _{0}}{\gamma -1}r^{-\left( 3\gamma -1\right) /2}\sin \left( \frac{%
r^{\gamma +1}}{\gamma +1}-\chi _{0}\right) ,  \label{asympt-2D}
\end{equation}%
and the TOC term is%
\begin{equation}
\mathrm{TOC}\left( \varphi _{\mathrm{asympt}}^{\mathrm{(2D)}}(x)\right)
=\varphi _{0}\frac{\left( \gamma +1\right) ^{2}-4S^{2}}{2\left( 2\gamma
+1\right) }r^{-3(\gamma +1)/2}\sin \left( \frac{r^{\gamma +1}}{\gamma +1}%
-\chi _{0}\right) ,  \label{TOC-2D}
\end{equation}%
cf. its 1D counterpart (\ref{TOC}). As above, $\varphi _{0}$ and $\chi _{0}$
are indefinite constants, in the framework of the asymptotic expansion at $%
r\rightarrow \infty $. Note that only the higher-order correction to the
asymptotic expansion, given by Eq. (\ref{TOC-2D}), includes the vorticity.

As seen from expression (\ref{asympt-2D}), the 2D norm of these eigenstates,
$N_{\mathrm{2D}}=2\pi \int_{0}^{\infty }$ $\varphi ^{2}(r)rdr$, converges
under precisely the same condition as in 1D, \textit{viz}., $\gamma >1$,
i.e., if the 2D expulsive potential is \emph{steeper} than the isotropic
anti-HO (quartic) potential. A qualitative explanation for this
counter-intuitive conclusion is the same as in the 1D setting: under the
action of the expulsive potential, a classical particle would be rolling
down the steep potential hill (along a spiral trajectory, if the particle
carries the angular momentum, corresponding to $S\geq 1$ in expression (\ref%
{psiphi})), with rapid acceleration. In terms of the particle's quantum
counterpart, the acceleration gives rise to the strongly oscillating phase
of the wave function, thus leading to the effective \emph{linear} \emph{%
self-trapping} (localization) of the eigenstate, with the normalizable
wavefunction.

Similar to the 1D case, for the 2D Schr\"{o}dinger equation (\ref{2DNLS})
with $\gamma =1$ (the anti-HO expulsive potential), the asymptotic
expression (\ref{asympt-2D}) takes a different form, with the logarithmic
correction to the phase of the rapid oscillations:%
\begin{equation}
\varphi _{\mathrm{asympt}}^{\mathrm{(2D)}}(r;\gamma =1)=\varphi
_{0}r^{-1}\cos \left( \frac{r^{2}}{2}+E\ln \left( \frac{r}{l}\right) \right)
,  \label{gamma=1 2D}
\end{equation}%
where\ $l$ is the radius of the inner core, cf. Eq. (\ref{gamma=1}). Also
similar to its 1D counterpart, in the case of $\gamma =1$ the norm of the 2D
eigenstate is weakly divergent, $N_{\mathrm{2D}}\simeq \pi \varphi
_{0}^{2}\ln \left( L/l\right) $ (cf. Eq. (\ref{N})), where $L$ is the radius
of the integration domain.

\begin{figure}[h]
\begin{center}
\includegraphics[height=5.0cm]{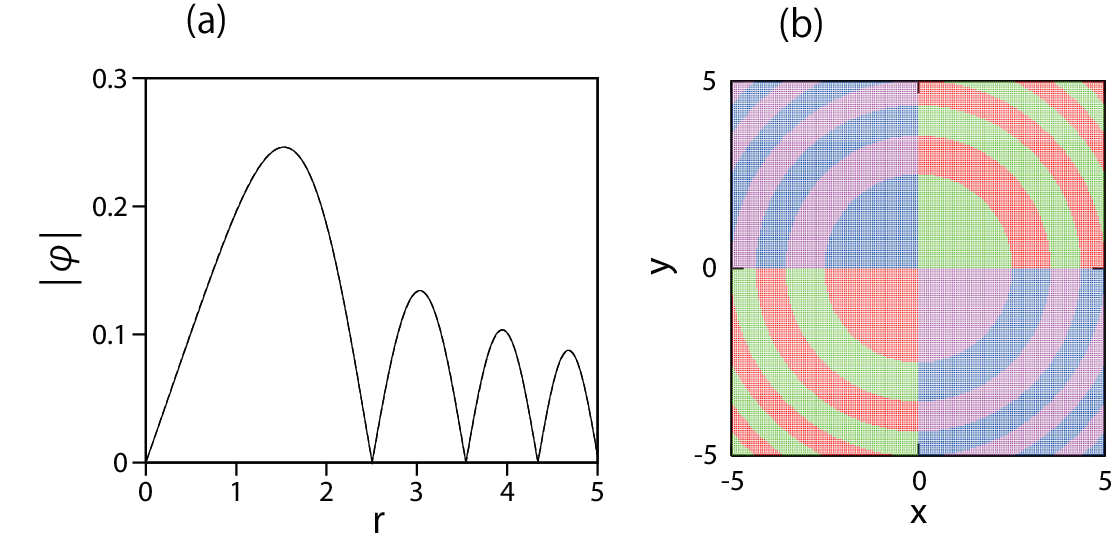}
\end{center}
\caption{(a) The absolute value $\left\vert \protect\varphi (r)\right\vert $
of the exact solution (\protect\ref{exact}) of Eq. (\protect\ref{g}) with $%
\protect\varphi _{0}=0.409$ for $\protect\gamma =1$, $g=0,$ $S=1$, and $E=0$%
. (b) The color snapshot of this 2D vortex state at $t=0$, $\protect\psi %
\left( r,\protect\theta \right) \equiv \protect\varphi (r)e^{i\protect\theta %
}$: Re$\protect\psi >0$ and Im$\protect\psi >0$ in the green region, Re$%
\protect\psi <0$ and Im$\protect\psi >0$ in the blue region, Re$\protect\psi %
<0$ and Im$\protect\psi <0$ in the red region, Re$\protect\psi >0$ and Im$%
\protect\psi <0$ in the purple one.}
\label{fig9}
\end{figure}

An additional analytical result, which is not available in the 1D case, is
the existence of particular \emph{exact solutions} to the linear version Eq.
(\ref{g}), with $g=0$ and $E=0$, if the power factor $\gamma $ of the
expulsive potential in Eq. (\ref{2DNLS}) is selected, for given integer
vorticity $S\geq 1$, as%
\begin{equation}
\gamma (S)=2S-1.  \label{gamma(S)}
\end{equation}%
In this case, the exact wave function is%
\begin{equation}
\varphi _{\mathrm{exact}}(r;S)=\frac{\varphi _{0}}{r^{S}}\sin \left( \frac{%
r^{2S}}{2S}\right) ,  \label{exact2}
\end{equation}%
where $\varphi _{0}$ is an arbitrary amplitude. For all values $S\geq 2$,
the norm of the exact wave function (\ref{exact2}) converges, yielding
\begin{equation}
\left( N_{\mathrm{2D}}\right) _{\mathrm{exact}}=2\pi \int_{0}^{\infty
}\varphi _{\mathrm{exact}}^{2}(r;S)rdr=\varphi _{0}^{2}\frac{\pi \Gamma
\left( 1/S\right) }{2S^{1-(1/S)}\left( S-1\right) }\cos \left( \frac{\pi }{2}%
\left( 1-\frac{1}{S}\right) \right) .  \label{S}
\end{equation}%
where $\Gamma \left( 1/S\right) $ is the Gamma-function. In agreement with
the above conclusions, the norm diverges for $\gamma =1$, i.e., $S=1$, see
Eq. (\ref{gamma(S)}), the respective exact solution (\ref{exact2}) taking
the form of%
\begin{equation}
\varphi _{\mathrm{exact}}(r;S=1)=\frac{\varphi _{0}}{r}\sin \left( \frac{%
r^{2}}{2}\right) .  \label{exact}
\end{equation}%
This analytical solution and the corresponding numerical solution of Eq.~(%
\ref{g}), with $\varphi _{0}=0.409$, are plotted by the solid and dashed
lines, respectively, in Fig. \ref{fig9}(a), where they completely overlap
(actually, the coincidence with the exact analytical solution corroborates
the accuracy of the numerical solution). The 2D structure of this vortex
eigenstate is illustrated by means of the color plot in Fig. \ref{fig9}(b).

Lastly, if the cubic term is retained in Eq. (\ref{g}) with $\gamma =1$ (the
anti-HO expulsive potential), one can construct an approximate solution for $%
E=0$ and $S\neq 1$, neglecting the third harmonic in the elementary formula $%
\sin ^{3}\Phi =(3/4)\sin \Phi -(1/4)\sin \left( 3\Phi \right) $. The
respective approximate solution is given by Eq. (\ref{exact}) (even if $S$
is not $1$), in which the amplitude is determined by the balance of the
linear and nonlinear terms in Eq. (\ref{g}):%
\begin{equation}
\varphi _{0}^{2}=2\left( 3g\right) ^{-1}\left( 1-S^{2}\right) .
\label{squared}
\end{equation}%
Thus, the approximate solution (\ref{exact}) of Eq. (\ref{g}) with the
self-defocusing or focusing nonlinearity, i.e., $g=+1$ or $g=-1$, exists,
yielding $\varphi _{0}^{2}>0$ as per Eq. (\ref{squared}), for $S=0$ or $%
S\geq 2$, respectively.

In the case of $g=-1$ (self-focusing), it is necessary to test the stability
of the nonlinear vortex solutions against spontaneous splitting, cf. Refs.
\cite{AB,CC,DD}. These results will be reported elsewhere.

\section{Conclusions}

Our analysis demonstrates that, on the contrary to\ the intuitive
expectation, the 1D and 2D Schr\"{o}dinger equations with the expulsive
potential which is steeper than the quadratic (anti-HO) one, generates a
full continuous spectrum of the \emph{normalizable} bound states, which may
be considered as a manifestation of the \emph{effective self-trapping in the
linear system}. In 1D, these bound states may be the fundamental and dipole
ones (spatially even and odd modes, respectively; plausibly, higher-order
excited states exist too, which is a subject for additional analysis). The
2D bound states may carry any integer vorticity (angular momentum).
Universal asymptotic expressions for the 1D and 2D states, valid at $%
|x|\rightarrow \infty $ and $r\rightarrow \infty $, respectively, are found,
providing an extremely accurate approximation to the numerically found wave
functions. In addition, in the 2D case special vorticity-carrying solutions
are found in the exact form. These results essentially extend the concept of
BIC (bound states in the continuum), which has recently drawn much interest
in the context of photonics, where the paraxial-propagation models amount to
the same Schr\"{o}dinger equations as in quantum mechanics. In the case of
the anti-HO (quadratic) shape of the expulsive potential, the normalization
integrals for the bound states are weakly divergent, in 1D and 2D settings
alike. The nonlinear version of the 1D and 2D models, in the form of the
respective GPEs (Gross-Pitaevskii equations), are briefly considered too. In
1D, the cubic self-focusing or defocusing nonlinearity slightly deforms the
linear solutions. A parity-breaking instability occurs in the case of the
self-focusing nonlinearity if the norm of the spatially even eigenstate
exceeds a critical value.
%The quintic self-focusing term can make the 1D bound states unstable against the onset
%of the collapse if the norm of the state exceeds the critical value.
A detailed analysis of the nonlinear models will be reported elsewhere. In
particular, a challenging issue is the (in)stability of 2D nonlinear vortex
eigenstates against splitting\ by the azimuthal modulational instability in
the case of the self-focusing cubic nonlinearity.~In parallel, it is
relevant to perform the spectral stability analysis of the nonlinear
eigenstates, in the framework of the Bogoliubov-de Gennes equations for
small perturbations~\cite{BdG,tkkp}.~As concerns the development of the
numerical analysis of the 1D and 2D eigenstates reported in this work, a
challenging issue is the need of using a large number of grid points for
capturing rapid oscillations of the decaying tails.~A potential way to
improve this aspect of the 2D solutions may be the use of the mesh
adaptation \cite{egc44}, that should help resolving the oscillations very
accurately with less computational cost.~These studies are currently
underway and will be reported elsewhere.

\section{Acknowledgments}

We thank P. G.~Kevrekidis and D. H. J. O'Dell for helpful discussions.

\section{Funding}

E.G.C. acknowledges support from the San Diego State University, Department of
Mathematics and Statistics startup fund.

\section{Author Contributions}

Software development and numerical calculations: H.S., A.C.A., and E.C.
Analytical considerations: B.A.M. and H.S. Analysis of the results: all
authors. Drafting the manuscript: B.A.M. H.S, and A.C.A.

\section{Conflict of Interests}

Authors declare no conflict of interests

\section{Data Availability Statement}

The data supporting the findings of this publication can be made available
upon request.

\end{document}